\documentclass[epjCONF, onecolumn]{svjour}
\usepackage{graphicx}
\usepackage[varg]{txfonts} 
\usepackage[latin1]{inputenc}
\session-title{Assembling the Puzzle of the Milky Way}

\newcommand{\fig}{Fig.~\ref}

\begin{document}
\title{On the age of field halo stars}
\author{Paula Jofr\'e\inst{1,2}\fnmsep\thanks{\email{jofre@obs.u-bordeaux1.fr}}}
\institute{Max Planck Institute for Astrophysics \and Laboratoire d'Astrophysique de Bordeaux}
\abstract{
A study of stellar ages of a sample from the Sloan Digital Sky Survey (SDSS) is presented. The results are consolidated with a set of globular clusters (GCs) and show that this stellar sample is composed by one dominant population of 10-12 Gyr old. This supports the Eggen's scenario claiming that the inner halo of the Milky Way formed rapidly, probably during the collapse of the proto-Galactic cloud. 
} 
\maketitle
\section{Introduction}
\label{intro}
There are two main formation scenarios for the Galactic halo. The first one \cite{sz} states that the stars in the halo formed slowly and chaotically via merging processes and the second one \cite{eggen} proposes a rapid formation during the collapse of the proto-Galactic cloud. In the first scenario, the halo is composed by stellar populations of different chemical abundances, kinematics and ages. The product of this chaotic formation has been observed via different evidence \cite{streams,watkins,tolstoy}. In the second formation scenario, the halo is composed by coeval stellar populations, which is supported through studies of globular clusters (GCs) \cite{SW02}. To distinguish between these scenarios and to understand how the Galactic halo might have formed the ages of field stars play a crucial role. In this report, I present how the ages of field stars can be determined. 

\section{Main Sequence Turn-Off}

A classical way to determine the age of a star is using isochrones. One has to place the temperature (or color) and the luminosity of the star in the Hertzsprung-Russell Diagram (HRD) and to find the isochrone with the corresponding metal content that crosses this region in the HRD. The determination of the age of an individual halo star is, however, very uncertain because it is almost impossible to place the star in a restricted region of the HRD. Firstly, the measurements of atmospheric parameters are uncertain, e.g.  an error of 100~K in the temperature can produce an error of up to 2~Gyr in the age determination of population II stars. Secondly,  the distance and the mass of halo stars are usually not known, which does not allow to know accurately their intrinsic luminosity. 

The age of halo stars can be studied by considering whole groups of stars. For coeval stars at a given metallicity,  the bluest ones are those at the turn-off point. The age of this population can then be determined by matching the turn-off temperature to the corresponding isochrone.  This study requires independent measurements of metallicities and temperatures of a large sample of stars, from where the turn-off can be determined. 

The sample was taken from the Sloan Digital Sky Survey \cite{sdss}. The selection of halo stars was done by using color-color diagrams to focus on the F-G dwarfs and subgiants.  The final sample contained approximately 100,000 stars with low-resolution spectra, from which the metallicity and temperature was estimated. This was done by performing a fitting between the observed spectra and a library of synthetic spectra \cite{max,jofre_age}.

\begin{figure}[t]
\centering
\includegraphics[scale=0.35,angle=90]{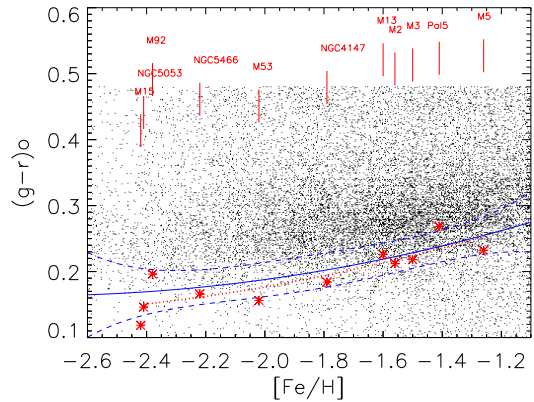}
\vspace{0.5cm}
 \caption{Metallicities and colors of a sample of field halo stars. Red asterisks are the turn-off color and metallicities of globular clusters observed with sloan colors. Blue line is correspond to the turn-off color as a function of metallicity  of the field population, with the dashed line representing the error. }
\label{fehteff}       
\end{figure}

In a coeval stellar population of a given metallicity, the main-sequence turn-off  shows an edge in the temperature distribution function. Figure~\ref{fehteff} shows the metallicity and the $(g-r)$ color of the SDSS stellar sample used for this study. While  the number of stars increases smoothly at red colors, at the blue there is an abrupt decrease, which is shown with the solid line. There is a significant number of stars bluer than the value given by this line. These stars can be blue horizontal branch stars, stars of the same age but metal-poorer, blue stragglers, or younger stars of the same metallicity.  
It is also important to mention that an older population, which would have a colder turn-off color than the dominating population, may well be hidden in this sample. Therefore, the turn-off represented by the blue edge can only estimate the age of the youngest dominant population. 

Technically, the turn-off detection was performed applying the Sobel Kernel \cite[and references therein]{Tabur09} to the temperature distribution of the stars at a given metallicity value. The main ingredient of this technique is a first derivative operator that computes the rate of change across an edge, where the largest change corresponds to the edge, i.e. the turn-off. For details on the turn-off detection and the error determination see \cite{jofre_age}. The blue line in \fig{fehteff} represents the result of the Sobel Kernel with the dashed line its error. 

\section{Age of the halo stars}
\begin{figure}[t]
\centering
\includegraphics[scale=0.5,angle=90]{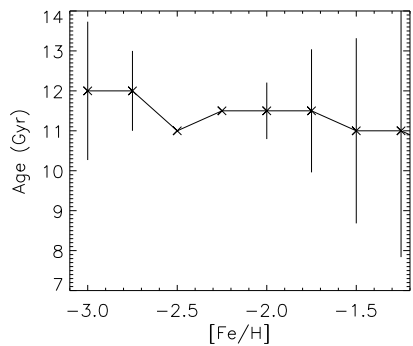}
 \caption{The age as a function of metallicity of the dominant halo-field population}
\label{age}       
\end{figure}

The turn-off temperature as a function of metallicity was used to look for the corresponding isochrone. It is important to remark that the absolute age determined only from the turn-off temperature is very sensitive to the effects of gravitational setting \cite{chaboyer92a,idea}. Here I show only the results obtained from isochrones which include atomic diffusion of the Garching Stellar Evolution Code \cite{garstec}. For an extensive discussion see \cite{jofre_age}.  The age as a function of metallicity is illustrated in Figure \ref{age}, were no significant age gradient is obtained when considering the errors. 

\subsection{Globular Clusters}
As tracers of the Galaxy, GCs can be treated as single stellar populations and their  ages are known with better confidence than those of field stars. For this reason, a sample of GCs with $ugriz$ photometry \cite{An08} was used to consolidate the results obtained for the field stars. 
The Sobel Kernel technique was also applied to the $(g-r)$ color distribution of the GC stars in the same way as for the field stars, whose blue edge is plotted in the diagram of Figure~\ref{fehteff} with asterisks. GC and field stars have same turn-off color as a function of metallicity, implying that absolute ages should be the same too. The study of the absolute age of 55 GCs \cite{SW02},  in which 7 clusters overlap with the sample of GCs of this study, conclude that they are 10-12 Gyr old, showing no gradient  of age as a function of metallicity.

%

Figure \ref{col_distr} shows the color distribution of an example GC  together with the color distribution of the field stars with the corresponding cluster metallicity. Guided by the color-magnitude diagram of the cluster,  the turn-off point is represented by the major peak in the distribution. The second minor peak towards red color is produced by the stars at the base of the red giant branch. Since both color distributions have similar shapes, and since globular clusters are composed by coeval stars, the field stars of the inner halo provided by SDSS must belong to one dominant  coeval population. Moreover, since  the turn-off color is the same for the field and the clusters, field and cluster stars must be coeval to each other.

\section{Discussion}

Since this work uses the independent measurements of spectroscopic temperatures and metallicities, the stellar sample is composed mostly by those belonging to the inner halo. The absence of a gradient as a function of metallicity found for this sample (see \fig{age}) hints that there is one dominant population of stars in the inner halo, where the star formation process took place 10-12 Gyr ago.

\begin{figure}[t]
\centering
\includegraphics[scale=0.45,angle=90]{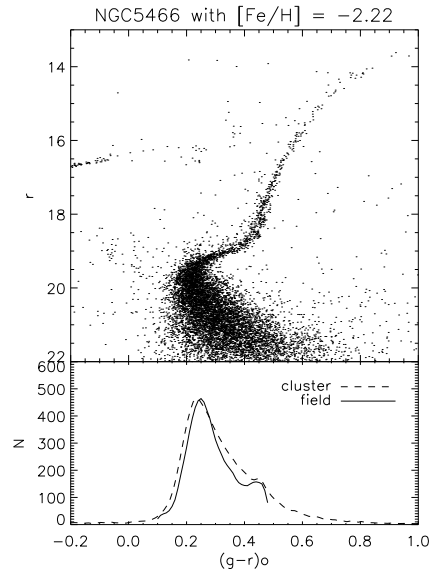}
 \caption{Upper panel: Color-magnitude diagram of an example cluster with $ugriz$ photometry. Lower panel: color distribution of the cluster stars (dashed line) compared with the color distribution of the field stars with the cluster metallicity (solid line)}
\label{col_distr}       
\end{figure}

In addition, there is empirical evidence that cluster and field stars share a similar history and are composed of a dominant population. The colors of sample of  clusters observed with SDSS \cite{An08} were compared to those of the field stars, where it was shown that both color distributions agreed especially in the turn-off color as a function of metallicity (see \fig{col_distr}).  The agreement between the ages for the field for GCs \cite{SW02}  serves as a further argument that the majority of inner halo stars are 10-12~Gyr old. \\

\noindent An old dominating population for the inner halo gives evidence of a  rapid formation of the halo during the collapse of the proto-Galactic gas. This scenario  agrees with conclusions from GC studies \cite{SW02,Sarajedini}, where the metal-poor clusters are coeval. Moreover, the absolute ages obtained for the clusters \cite{SW02} are also 10-12~Gyr, which implies that GCs and field stars are coeval to each other. 

In the stellar sample considered here there is a significant number of stars with bluer colors than the color cut-off of the main sequence turn-off (see \fig{fehteff}). They  have also  been noticed in similar works \cite[and references therein]{SN06}. An interesting explanation for them is that they were formed in  small external galaxies and have been accreted later on to the Milky Way halo. These galaxies  experience a different star formation history to the Milky Way and therefore can be younger than the dominating population of  field stars.  Further analysis of their kinematics and chemical abundances is needed to prove this scenario. These blue stars and the existence of a dominant population of halo stars suggest that the two rivaling formation scenarios of the Galactic halo \cite{sz,eggen} actually complete, in a composed manner, the picture of how the Milky Way might have formed. These scenarios combined propose that part of the halo has collapsed rapidly, while the other part  has been populated through collisions and mergers  between the satellite galaxies and our Milky Way.

%

%

\end{document}